\documentclass[conference]{IEEEtran}
\IEEEoverridecommandlockouts

\usepackage[T1]{fontenc}
\usepackage[utf8]{inputenc}
\usepackage{lmodern}

\usepackage{amsmath,amssymb,amsfonts}
\usepackage{amsthm}
\usepackage{mathtools}
\usepackage{bm}
\usepackage{bbm}          

\usepackage{graphicx}
\usepackage{booktabs}
\usepackage{array}
\usepackage{multirow}
\usepackage{colortbl}
\usepackage{wrapfig}
\usepackage{stfloats}     
\usepackage[caption=false,font=footnotesize]{subfig}
\usepackage{float}

\usepackage{microtype}
\usepackage{xcolor}
\usepackage{xspace}
\usepackage[shortlabels]{enumitem}
\usepackage{textcomp}     
\usepackage{verbatim}
\usepackage{fancybox}
\usepackage{multicol}
\usepackage{todonotes}

\usepackage{url}
\usepackage[hidelinks]{hyperref}
\usepackage[capitalise,noabbrev,nameinlink]{cleveref}  
\usepackage{cite}          

\usepackage[ruled, noline, linesnumbered, commentsnumbered, noend]{algorithm2e}

\usepackage{balance}  

\allowdisplaybreaks
\usepackage{url}

\usepackage{amsmath,amsfonts}

\newfont{\bbb}{msbm10 scaled 700}

\DeclareSymbolFont{bbsymbol}{U}{bbold}{m}{n}

\newfont{\bb}{msbm10 scaled 1100}

\DeclareMathSymbol{\CC}{\mathbin}{bbsymbol}{'103}
\DeclareMathSymbol{\PP}{\mathbin}{bbsymbol}{'120}
\DeclareMathSymbol{\RR}{\mathbin}{bbsymbol}{'122}
\DeclareMathSymbol{\QQ}{\mathbin}{bbsymbol}{'121}
\DeclareMathSymbol{\ZZ}{\mathbin}{bbsymbol}{'132}
\DeclareMathSymbol{\FF}{\mathbin}{bbsymbol}{'106}
\DeclareMathSymbol{\GG}{\mathbin}{bbsymbol}{'107}
\DeclareMathSymbol{\EE}{\mathbin}{bbsymbol}{'105}
\DeclareMathSymbol{\NN}{\mathbin}{bbsymbol}{'116}
\DeclareMathSymbol{\KK}{\mathbin}{bbsymbol}{'113}
\DeclareMathSymbol{\HH}{\mathbin}{bbsymbol}{'110}
\DeclareMathSymbol{\SSS}{\mathbin}{bbsymbol}{'123}
\DeclareMathSymbol{\UU}{\mathbin}{bbsymbol}{'125}
\DeclareMathSymbol{\VV}{\mathbin}{bbsymbol}{'126}
\DeclareMathSymbol{\XX}{\mathbin}{bbsymbol}{'130}
\DeclareMathSymbol{\BB}{\mathbin}{bbsymbol}{'102}

\DeclareMathSymbol{\yy}{\mathbin}{bbsymbol}{'171}
\DeclareMathSymbol{\xx}{\mathbin}{bbsymbol}{'170}
\DeclareMathSymbol{\zz}{\mathbin}{bbsymbol}{'172}
\DeclareMathSymbol{\sss}{\mathbin}{bbsymbol}{'163}
\DeclareMathSymbol{\rr}{\mathbin}{bbsymbol}{'162}
\DeclareMathSymbol{\pp}{\mathbin}{bbsymbol}{'160}
\DeclareMathSymbol{\qq}{\mathbin}{bbsymbol}{'161}
\DeclareMathSymbol{\ww}{\mathbin}{bbsymbol}{'167}
\DeclareMathSymbol{\hh}{\mathbin}{bbsymbol}{'150}
\DeclareMathSymbol{\uu}{\mathbin}{bbsymbol}{'165}
\DeclareMathSymbol{\vvv}{\mathbin}{bbsymbol}{'166}
\DeclareMathSymbol{\ee}{\mathbin}{bbsymbol}{'145}

\usepackage[mathscr]{euscript}

\newcommand{\hv}{{\bf h}}

\newcommand{\uv}{{\bf u}}

\newcommand{\vvec}{{\bf v}}

\newcommand{\zerov}{{\bf 0}}

\newcommand{\Dm}{{\bf D}}

\newcommand\varcal[1]{\text{\usefont{OMS}{cmsy}{m}{n}#1}}

\newcommand{\Cc}{\varcal{C}}

\newcommand{\Kc}{\varcal{K}}
\newcommand{\Lc}{\varcal{L}}
\newcommand{\Nc}{\varcal{N}}

\newcommand{\Qc}{\varcal{Q}}

\newcommand{\Uc}{\varcal{U}}

\newcommand{\Sigmam}{\hbox{\boldmath$\Sigma$}}

\newcommand{\trace}{{\hbox{tr}}}

\newcommand{\eqdef}{\stackrel{\Delta}{=}}

\newcommand{\herm}{{\sf H}}

\newcommand{\transp}{{\sf T}}

\newcommand{\SINR}{{\sf SINR}}
\newcommand{\SNR}{{\sf SNR}}

\newcommand{\fhratekUL}{B_{\ell, k}}
\newcommand{\fhratekDL}{R^{\rm dl}_k}
\newcommand{\demmatrix}{\Dm_F}

\usepackage{xparse}

\newcommand{\defeq}{ \stackrel{\triangle}{=} }

\definecolor{arashcolor}{rgb}{0.0735, 0.29, 0.78}

\def\BibTeX{{\rm B\kern-.05em{\sc i\kern-.025em b}\kern-.08em
    T\kern-.1667em\lower.7ex\hbox{E}\kern-.125emX}}
\begin{document}

\title{Rethinking Fronthaul Topologies \\for Cell-Free 6G Networks
}

\author{
     Max Franke\IEEEauthorrefmark{1},
     Arash Pourdamghani\IEEEauthorrefmark{1},
     	\IEEEauthorblockN{Fabian G\"{o}ttsch\IEEEauthorrefmark{1}\IEEEauthorrefmark{2}, Stefan Schmid\IEEEauthorrefmark{1}
        and
            Giuseppe Caire\IEEEauthorrefmark{1}\IEEEauthorrefmark{2} 
	}
\IEEEauthorblockA{\IEEEauthorrefmark{1}TU Berlin, \IEEEauthorrefmark{2}Massive Beams}
}
\maketitle

\begin{abstract}



Due to significant progress in physical layer (PHY) technologies, future 6G networks are expected to feature much denser antenna deployments. Cell-free MIMO network designs are one of the most promising candidates to enable denser networks. While significant effort has been put into its PHY research, the challenges it brings to networking disciplines remain relatively unexplored. In this work, we propose various fronthaul network designs for cell-free MIMO. Our results show that while tree topologies may suffice for small-scale deployments, they become infeasible as the number of antennas increases. In contrast, with growing network size, the proposed Clos topology performs almost as well as the optimal topology.
\end{abstract}

\begin{IEEEkeywords}
Fronthaul, Cell-Free MIMO, Topologies, Optimization
\end{IEEEkeywords}

\section{Introduction}

\begin{figure}[h]
  \centering
  \doublebox{
    \begin{minipage}{0.4\textwidth} 
      \centering
      \subfloat[Distributed MIMO]{%
        \includegraphics[width=0.45\linewidth]{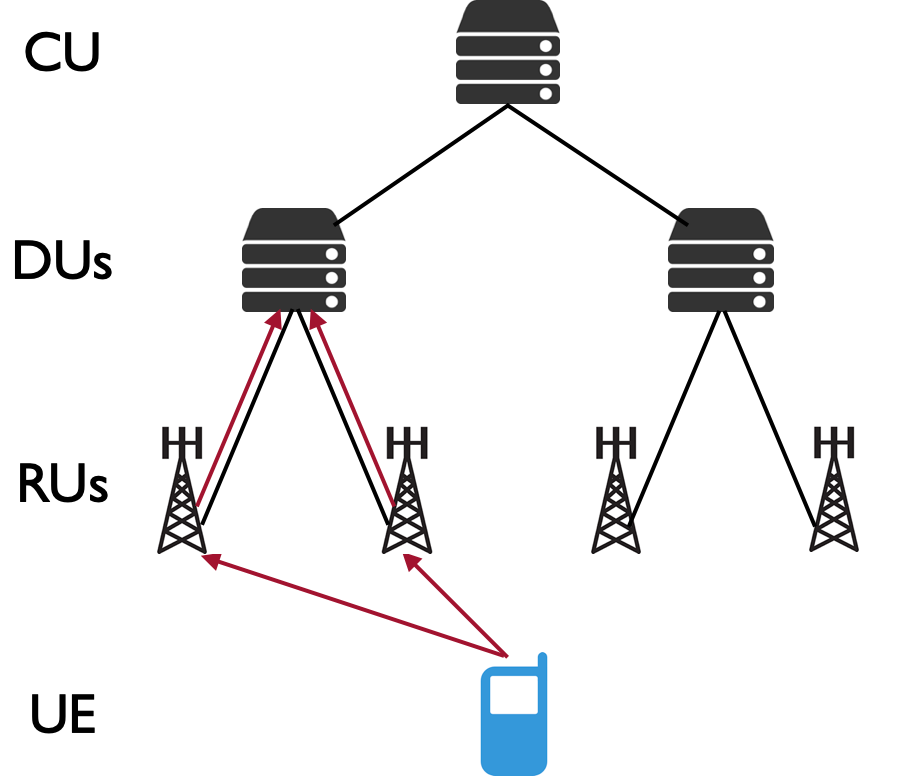}%
        \label{fig:SoTA}%
      }\hfill
      \subfloat[Cell-free MIMO]{%
        \includegraphics[width=0.45\linewidth]{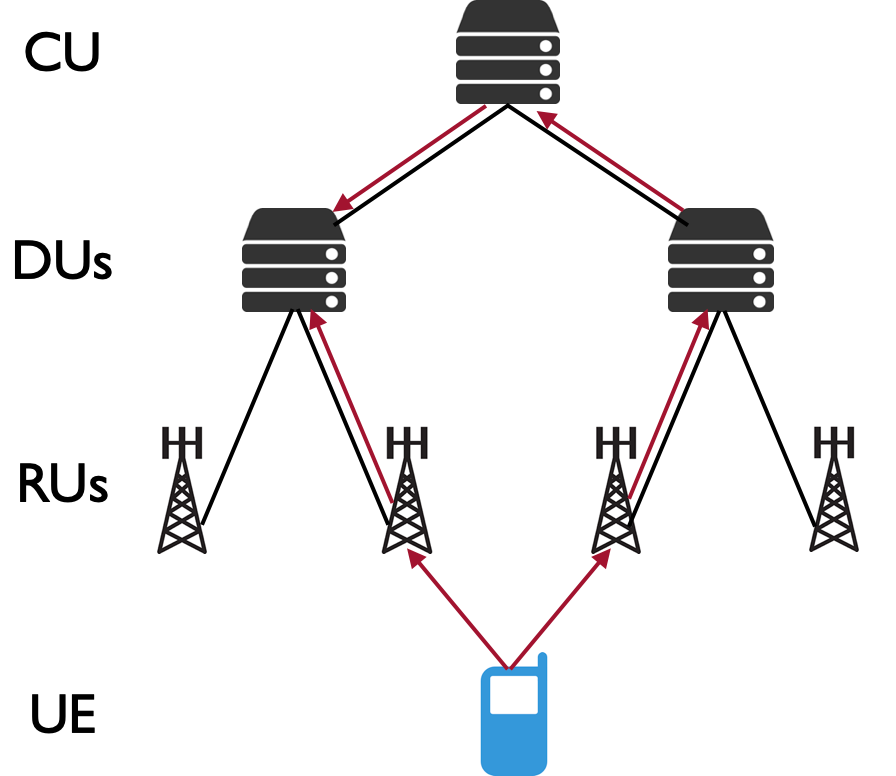}%
        \label{fig:CF}%
      }\\[5pt] 

      \subfloat[Inter DU]{%
        \includegraphics[width=0.45\linewidth]{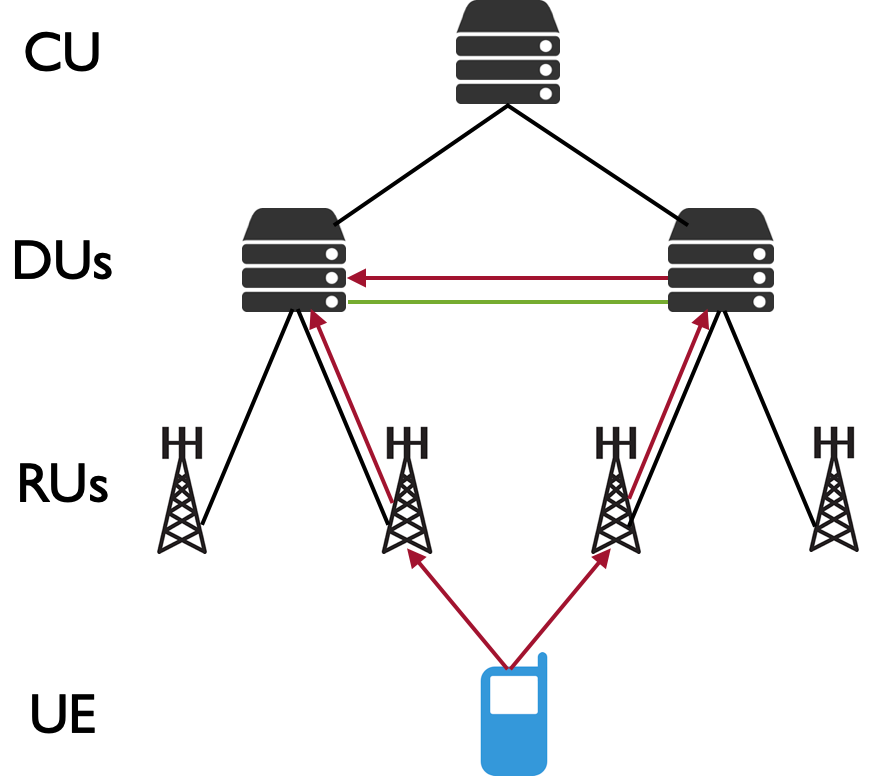}%
        \label{fig:iDU}%
      }\hfill
      \subfloat[Routed Fronthaul]{%
        \includegraphics[width=0.45\linewidth]{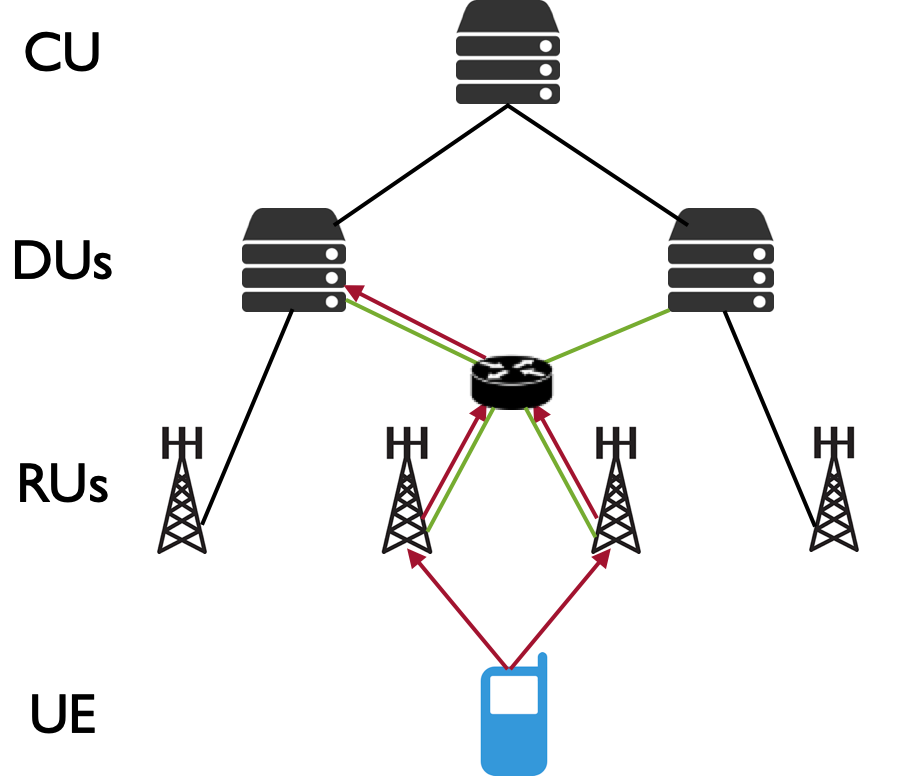}%
        \label{fig:rFH}%
      }%
    \end{minipage}%
  }%
  \label{fig:topos}
  \caption{
  This figure depicts different radio access and fronthaul network designs. 
  }
  \label{fig:grid}
\end{figure}

The wireless communications community has long searched for new methods to tackle inter-cell interference. One approach is to distribute the antennas of a BS over the cell area, a so-called distributed MIMO system. A suitable architecture for distributed MIMO is proposed by the O-RAN alliance, where the monolithic BS is separated into antenna sites, i.e., radio units (RUs), and a distributed unit (DU). The RUs handle the radio frequency (RF) processing
and are typically located close to the antennas, while the DU acts as processing node responsible for network resource allocation and radio link control. 
In a distributed MIMO system with network-centric clusters, a UE can be connected to all RUs associated with the DU in a given cell, allowing for increased macrodiversity.
Hence, even if the UE is located close to the cell-edge, it will maintain good channel gains with some of the RUs, such that the quality of service inside the cell is distributed more uniformly.


In an ideal \textit{user-centric} cell-free network, each UE is connected to all surrounding RUs with good channel characteristics. Signals from other RUs not serving the UE are then typically weak and do not cause significant interference. This user-centric formation of RU clusters 
is the main difference and advantage compared to cell-based systems. The advantages of user-centric cell-free MIMO compared to cellular MIMO and small cell networks have been pointed out in several works \cite{ngo2024ultradense, DemirFoundation}. 

While user-centric cell-free MIMO obtains impressive results in the physical layer, it also introduces new challenges to the higher layers in terms of architecture and protocol design. In current deployments, DUs are only indirectly connected through centralized units (CUs), which sit one layer above the DUs. 
In today’s O-RAN deployments, DUs communicate only via a centralized CU, which limits scalability (Fig. \ref{fig:CF}).
To lower fronthaul load, we envision some local and cluster-level processing, meaning that some functions, such as channel estimation, are carried out at the RUs directly. Additionally, we propose a routed fronthaul network between RUs and DUs (as shown in Figs. \ref{fig:iDU} and \ref{fig:rFH}), avoiding CU bottlenecks.
\subsection{Related Work}
While the physical layer of cell-free networks has been extensively investigated in recent years (see \cite{ngo2024ultradense} and references therein), the fronthaul is still very little studied. The vast majority of earlier works consider a centralized-RAN-type network, where the cluster-level processing of all UEs is done at a centralized unit \cite{Bashar_2018, masoumi2019performance}, which results in a non-scalable network.
Only recently, a few works have been published that study the cluster processor placement at one of multiple DUs \cite{li2023joint, joshi2024fronthaul, lai2024hybrid}.
In \cite{lai2024hybrid}, the objective of the proposed problem is to jointly minimize the fronthaul load and maximize the physical layer data rate of all UEs by optimizing the UE-RU associations.
A routed fronthaul network is considered in \cite{li2023joint, joshi2024fronthaul, lai2024hybrid}, where routing is done via dedicated routers placed in the fronthaul network \cite{li2023joint} or routers co-located at DUs, utilizing inter-DU links \cite{joshi2024fronthaul, lai2024hybrid}.


From an algorithmic point of view, our contribution lies in the intersection of demand-aware network design~\cite{avin2018toward} and traffic engineering~\cite{Weight2}.
In the context of network design, moving beyond simple topologies such as trees, while considering the potential costs of more complex configurations, has been examined in various settings, including data center networks~\cite{Al-Fares08Fat-Tree,SeedTree}.
One of the most well-known examples is Clos topology~\cite{clos1953study} used in large-scale datacenter networks~\cite{ClosGoogle, ClosMicrosoft} and adopted in this work.
From traffic engineering point of view, prior work mostly either focused solely on unicast~\cite{SDN2,SDN1} or multicast~\cite{rost2013virtucast}. However, since DL traffic utilizes multicast, our model in this paper needs to support both unicast and multicast in the same instance.
\subsection{Our Contributions}
In this paper, we propose different fronthaul topologies and study their impact on fronthaul performance in terms of link load. To this end, we develop a realistic model for traffic demands and explore load optimization opportunities. Following this, we explore and contribute efficient algorithms for processing node assignment, highlighting the effect of the interplay between node assignment and topology. We then empirically evaluate our optimizations in different scenarios.
The results give an insight into the optimal design of the fronthaul network architecture and show the impact of the processing node assignment algorithm on different metrics, such as maximum link traffic. While previous work has investigated different fronthaul topology desgins\cite{furtado2022cell}, this was done mostly from a technoeconomic standpoint.

Our main findings are the following.
\begin{enumerate}
    \item Tree topologies, the current state-of-the-art solution, become unfeasible as the network size increases. In contrast, the performance of the proposed Clos topology approaches that of the optimal topology for large networks that we created using an MILP. Interestingly, the Clos topology results in maximum link loads of around 200 Gbps only, meaning off-the-shelf hardware can handle these demands.
    \item Fronthaul traffic demands are predicted to increase further with the introduction of higher frequency bands, such as mmWave and sub-THz. As tree topologies are already reaching the limit of available link capacities, smarter fronthaul network designs, like the Clos topology, should be investigated further.
    \item The choice of algorithm that assigns UEs to DUs can play a significant role, offering fronthaul traffic reductions of up to 20\%. 
\end{enumerate}
\label{sec:background}
\begin{figure}[t!]
    \centerline{\includegraphics[trim={0 0 0 0}, clip,width=0.8\linewidth]{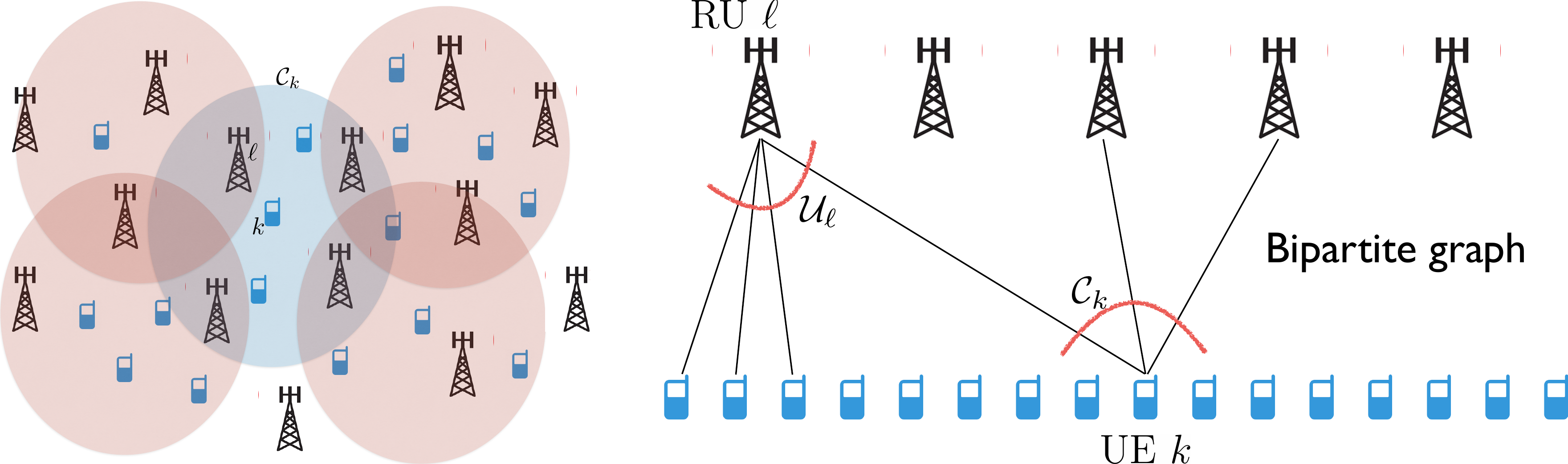}}
    \vspace{-.0cm}
    \caption{An example of user-centric clusters and UE-RU association. The sets $\Uc_\ell$ of UEs associated to each RU $\ell$ and the sets $\Cc_k$ of RU serving UE $k$ define a bipartite graph with RU and UE vertices and edges $(k,\ell)$ for all $k$ and $\ell$ such that $k \in \Uc_\ell$ and $\ell \in \Cc_k$.}
    \vspace{-.0cm}		
    \label{clusters}
\end{figure}

\section{Modeling Fronthaul Traffic} \label{sec:sys}
We consider a user-centric cell-free MIMO network consisting of the radio access network (RAN), i.e., the connections between the UEs and RUs, and a routed fronthaul network to enable the cluster-level signal processing at the DUs. 
 The network consists of $K$ single-antenna UEs, $L$ RUs, $Q$ routers, and $N$ DUs. 
Each RU is equipped with $M$ antennas, and the routers are located between RUs and DUs in the fronthaul network.
The sets of UEs, RUs, routers, and DUs are denoted by $\Kc=\{1,\dots, K\}$, $\Lc=\{1,\dots, L\}$, $\Qc=\{1,\dots, Q\}$, and $\Nc=\{1,\dots, N\}$, respectively. We follow closely the RAN model (and its implications for the fronthaul traffic) from \cite{li2023joint}, which will be briefly summarized in the following.

The wireless channel between RU $\ell$ and UE $k$ is given by $\hv_{\ell,k} \sim \mathcal{CN} ({\bf 0},\Sigmam_{\ell,k})$, i.e., a correlated complex circularly symmetric 
Gaussian vector with mean zero and 
covariance matrix 
$\Sigmam_{\ell,k} = \EE[ \hv_{\ell,k}\hv_{\ell,k}^\herm]$, 
where $\Sigmam_{\ell,k}$ describes the spatial characteristics of the wireless channel. 
The corresponding large-scale fading coefficient (LSFC) is 
    $\beta_{\ell,k} \defeq \frac{1}{M} \trace(\Sigmam_{\ell,k})$
and captures the average signal attenuation between 
RU $\ell$ and user $k$ due to distance and other macroscopic effects. 
The LSFCs are generated according to the 3GPP urban microcell street canyon pathloss model from \cite[Table 7.4.1-1]{3gpp38901}.
As shown in Fig. \ref{clusters}, each UE $k$ is associated to a user-centric cluster of surrounding RUs $\Cc_k \subseteq \Lc$ (and each RU $\ell$ serves a set of UEs $\Uc_\ell \subseteq \Kc$) and an UL pilot for channel estimation based on $\left\{\Sigmam_{\ell,k}, \beta_{\ell,k}\right\}$. This is done according to the cluster formation scheme in \cite{li2023joint}. 

Regarding the fronthaul, note that the user-centric cluster of RUs of each UE is associated to one of the DUs, which is responsible for the cluster-level processing. The DU is thus the \textit{cluster processor} of the user-centric cluster. In order to enable cluster-level processing, data must be exchanged between the RUs serving a given user and the corresponding cluster processor.

We use a standard block-fading model (e.g., see \cite{Noncooperative,Larsson-book,DemirFoundation} and references therein), where the UL and DL channel coefficients remain constant (channel reciprocity) over blocks of $T$ signal dimensions (time-frequency channel uses). 
We will refer to one block of $T$ signal dimensions with constant channel coefficients as ``coherence block'' in the following.
One coherence block consists of 
$\tau_p$ pilot symbols for channel estimation and $T - \tau_p$ payload data symbols. The channel estimate is obtained from the $\tau_p$ pilots using subspace-based channel estimation and used to compute the precoding and receiving vector \cite{Subspace-Based}. 
The network operates the UL and DL in time division duplex (TDD), such that disjoint fractions of the $T$ signal dimensions with constant channel coefficients are used for UL and DL, respectively.
	We denote by ${ {\gamma_{\rm DL}}} \in (0,1)$ the resource fraction allocated to the DL, and
	$(1 - { {\gamma_{\rm DL}}})$ is allocated to the UL. 

\subsection{Physical Layer Rates}


All UEs transmit with the same average energy per symbol $P^{\rm ue}$ and we define the system parameter 
$\SNR \defeq \frac{P^{\rm ue} }{N_0}$, where $N_0$ denotes the complex baseband noise power spectral density.
Following \cite{li2023joint}, we let $\alpha_{\ell ,k}$ and $\widehat{\sigma}^2_{\ell,k}$ the fronthaul quantization factor and error (to be defined later). The resulting UL Signal to Interference plus Noise Ratio (SINR) of user $k$ 
is given by \cite{li2023joint}
\begin{equation} 
    \SINR^{\rm ul}_k  = \frac{  \SNR \left |  \sum_{\ell \in \Cc_k} \widetilde{g}_{\ell, k, k} \right |^2 }
    {\sum_{\ell \in \Cc_k}  \widetilde{d}_{\ell,k}
    + \SNR \sum_{i \neq k} \left |  \sum_{\ell \in \Cc_k} \widetilde{g}_{\ell, k, i}  \right |^2 } ,  \nonumber
\end{equation}
where 
    $\widetilde{g}_{\ell, k, i} \eqdef w^*_{\ell,k} \alpha_{\ell,k} \vvec_{\ell,k}^\herm \hv_{\ell,i}$,
    $\widetilde{d}_{\ell,k} \eqdef |w_{\ell,k}|^2 \left ( \alpha_{\ell,k}^2 \|\vvec_{\ell,k}\|^2 + \widehat{\sigma}^2_{\ell,k} \right )$
and $w_{\ell,k}$ and $\vvec_{\ell, k}$ 
are the cluster-level combining coefficient\footnote{The cluster-level combining coefficients $w_{\ell,k}$ for all $(\ell,k)$ are chosen to maximize a nominal SINR expression of the channel given the local knowledge of the cluster processor $\Cc_k$ (see \cite{li2023joint} for details). 
Further, the coefficients are such that each $\vvv_k = \left[w_{1,k} \vvec_{1,k}^\transp \dots w_{L,k}\vvec_{L,k}^\transp \right]^\transp$ has unit norm.} and the receiving vector of $(\ell,k)$, respectively.
The receiving/combining vectors are computed based on a linear Minimum Mean-Square Error (LMMSE) principle as in \cite{li2023joint}.
As a performance measure of the UL physical layer,  we consider the so-called {\em Optimistic Ergodic Rate} (OER) \cite{8304782} given by
\begin{eqnarray}
    R^{\rm ul}_k = \EE [ \log (1 + \SINR^{\rm ul}_k) ], \label{ergodic_rate_ul}
\end{eqnarray}
where the expectation is with respect to small-scale fading.

For the downlink (DL) precoding vectors, we use the approximate UL-DL reciprocity of \cite{Subspace-Based} and let
$\uv_{\ell,k} \propto w^0_{\ell,k} \vvec_{\ell,k}$, where $w^0_{\ell,k}$ are the cluster-level combining coefficients for the case of zero quantization distortion, i.e., $\alpha_{\ell,k} = 1$ and $\widehat{\sigma}^2_{\ell,k} = 0$. 
The DL SINR is then given by 
\begin{eqnarray}
    \SINR^{\rm dl}_k & = & \frac{|\hh_k^\herm \uu_k|^2 q_k}{\SNR^{-1} + \sum_{j\neq k}   |\hh_k^\herm \uu_j|^2 q_j }, \label{DL-SINR} 
\end{eqnarray}
where $\hh_k = \left[\hv_{1,k}^\transp \dots \hv_{L,k}^\transp \right]^\transp$ and $\uu_k = \left[\uv_{1,k}^\transp \dots \uv_{L,k}^\transp \right]^\transp$. Note that $\uv_{\ell,k} \propto w^0_{\ell,k} \vvec_{\ell,k}$ for all $\ell \in \Cc_k$ and $\uv_{\ell,k} = \zerov$ if $\ell \notin \Cc_k$. Further, we normalize each $\uu_k$ to have unit norm.
For simplicity, in this paper we choose uniform power allocation to all data streams, i.e., $q_k = 1$ for all $k$ \cite{Subspace-Based}. 
The corresponding OER for the DL is given by 
\begin{eqnarray}
    R^{\rm dl}_k = \EE [ \log (1 + \SINR^{\rm dl}_k) ]. \label{ergodic_rate_dl}
\end{eqnarray}

\subsection{Fronthaul Network Load} \label{sec:fronthaul_load}
\begin{figure}
	\centerline{
		\includegraphics[width=0.38\linewidth, trim={279 208 253 122}, clip]{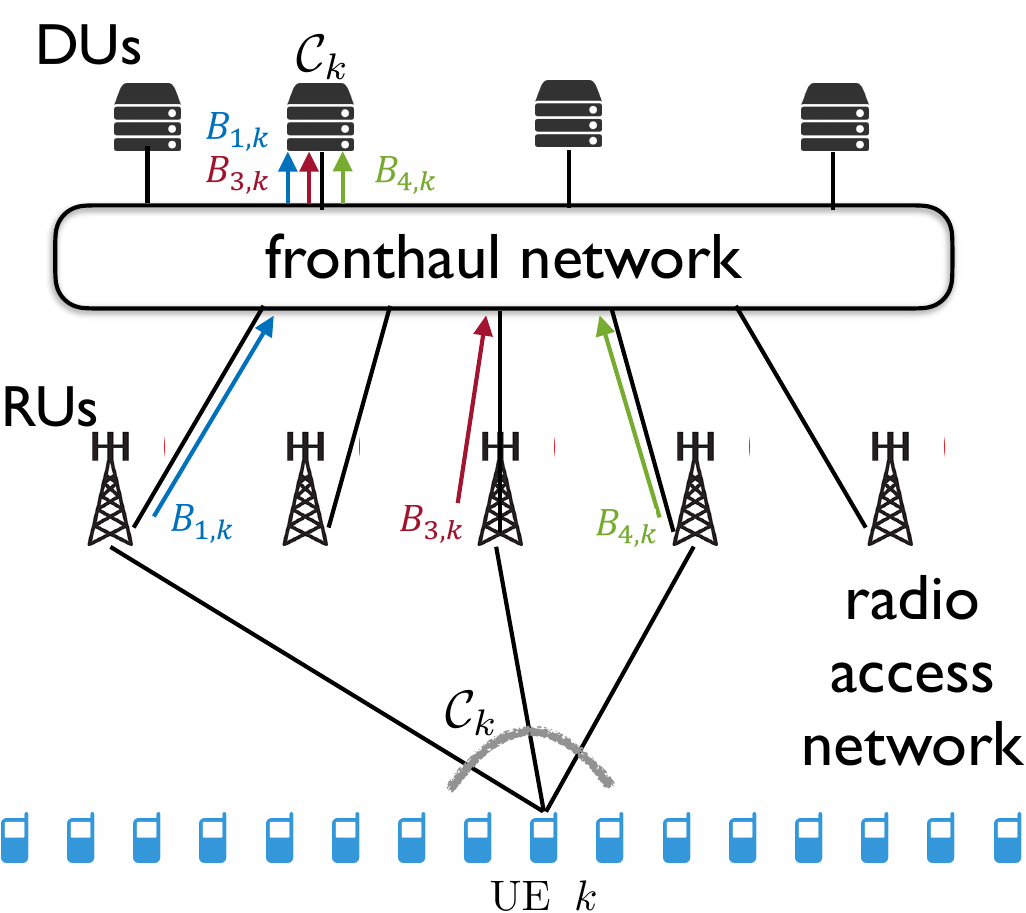} \hspace{.2cm}
		\includegraphics[width=0.38\linewidth, trim={279 208 253 122}, clip]{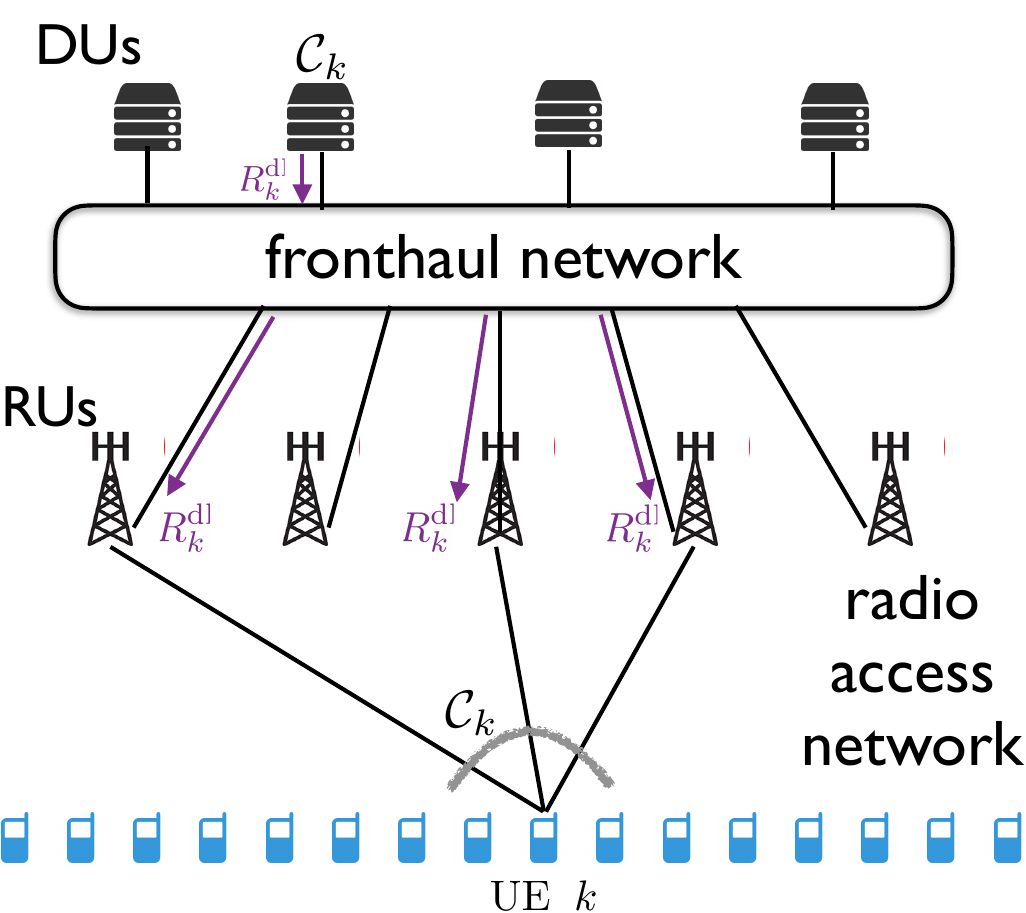}}
	\caption{Example of the fronthaul data exchange between RUs $\ell \in \Cc_k$ and the DU hosting $\Cc_k$ in the UL (left) and DL (right). 
    }
	\label{fig_fronthaul_flow_UL_DL}
\end{figure}

The fronthaul load in the UL for UE $k$ between a RU $\ell \in \Cc_k$ and the DU hosting $\Cc_k$ is given by 
\begin{equation} 
    B_{\ell,k}  = \max \left \{ \log_2 \frac{\sigma_{\ell,k}^2}{D}, 0 \right \},\label{rateQ} 
\end{equation}
where $\sigma_{\ell,k}^2$ and $D$ are the received signal strength after LMMSE combining at RU $\ell$ of UE $k$'s UL data signal and the quantization distortion level in the fronthaul, respectively \cite{li2023joint}. Note that these two quantities define the quantization factor $\alpha_{\ell,k} =  \frac{\sigma_{\ell,k}^2 - D}{\sigma_{\ell,k}^2}$
and error $\widehat{\sigma}^2_{\ell,k} = (1 - D/\sigma_{\ell,k}^2) D$. In the DL, all RUs in $\Cc_k$ transmit the same symbols to UE $k$. We assume that modulation and precoding are done at the RUs, such that the DU hosting cluster $\Cc_k$ sends the same information bits to the RUs in $\Cc_k$. 
The number of information bits per channel use necessary to encode the DL signal for user $k$ at each RU $\ell \in \Cc_k$ is equal to $R^{\rm dl}_k$ \cite{li2023joint}. The UL and DL fronthaul loads are depicted in Fig. \ref{fig_fronthaul_flow_UL_DL}.

\section{Modeling the Network}
\label{sec:Abstractions}

\subsection{Abstractions and Assumptions}
As mentioned, current state of the art for fronthaul networks are passive optical networks. These do not require any routing and are thus not susceptible to any issues related to it, such as delays caused by processing or queueing. However, since routing will be necessary in cell-free networks, we are considering only routed fronthaul networks in our evaluation. This means that at the very least, there will be a router co-located with each DU. Additional routers may also be added, depending on the given topology. It should be noted that it is beyond the scope of this paper to investigate how to facilitate such a routed fronthaul. 

As the processing delay is usually only in the range of a few microseconds \cite{angrisani2005measurement}, the more critical delay component added by routers is usually related to queueing. Such delay is highly dependent on metrics like congestion, buffer size etc. Emerging technologies such as optical switches can prevent queueing altogether and achieve submillisecond total switching times \cite{calvarese2022strategies}.  For these reasons, we choose to abstract delays and instead introduce a maximum hop count as a constraint to prevent unfeasibly deep networks. 

Regarding link capacities, PONs only reach 50 Gbps, with the next generation likely moving up to 100 Gbps~\cite{cwalina2023demonstration}.


\newcommand{\UE}{\textsc{UE}\xspace}
\newcommand{\subUE}{\textsc{SUE}\xspace}
\newcommand{\RU}{\textsc{RU}\xspace}
\newcommand{\DU}{\textsc{DU}\xspace}
\newcommand{\Router}{\textsc{R}\xspace}

\newcommand{\ru}
{\textsc{ru}\xspace}
\newcommand{\du}{\textsc{du}\xspace}
\newcommand{\router}{\textsc{r}\xspace}

\newcommand{\UEs}{\textsc{UEs}\xspace}
\newcommand{\subUEs}{\textsc{SUEs}\xspace}
\newcommand{\RUs}{\textsc{RUs}\xspace}
\newcommand{\DUs}{\textsc{DUs}\xspace}
\newcommand{\Routers}{\textsc{Rs}\xspace}

\definecolor{forestgreen}{rgb}{0.13, 0.55, 0.13}
\newcommand{\ctcp}[1]{\hfill \textcolor{forestgreen}{\# {#1}}}

\let\oldnl\nl
\newcommand{\nonl}{\renewcommand{\nl}{\let\nl\oldnl}}
\newcommand{\cfront}[1]{\nonl \textcolor{forestgreen}{\# {#1}}}

\SetNoFillComment 
\DontPrintSemicolon 
\SetKwFor{For}{for}{}{} 

\newcommand{\quotes}[1]{``#1''}
\subsection{Inputs \& Objective}
We consider the input to be in the form of $\langle N, Q, \demmatrix \rangle$ in which $N$ indicates number of \DUs, $Q$ indicates number of routers, and $\demmatrix$ is a \emph{fronthaul demand} matrix,  indicating fronthaul traffic (bandwidth) resulting from the physical layer communication between each associated \UE-\RU pair.
Furthermore, we assume a set of constraints for each network element. In particular $\Delta^{\DU}$, $\Delta^{\RU}$ and  $\Delta^{\Router}$, representing  the maximum number of connections of \DUs, \RUs and routers respectively. Additionally, we consider $\beta$ as the maximum number of \UEs that a \DU can handle.

Our algorithms aim to optimize for two metrics: (1) \emph{Link utilization} (the maximum amount of bandwidth that goes through a link, in either Uplink or Downlink direction), (2) \emph{Delay} (the maximum number of hops between \UEs and their assigned \DUs).
\subsection{Topology Design}
\label{sec:GraphModel}

\begin{figure*}[t]
    \centering
    \subfloat[Complete topology]{%
        \includegraphics[width=0.30\textwidth]{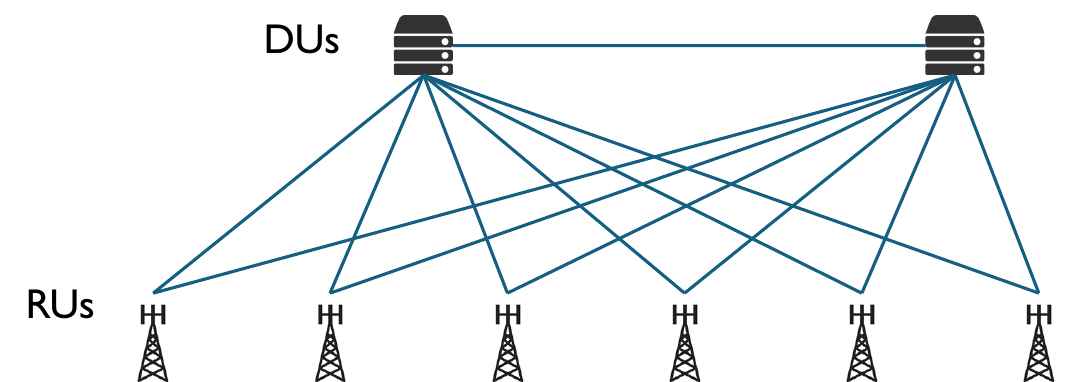}%
        \label{fig:complete-topology}%
    }\hfill
    \subfloat[Tree topology]{%
        \includegraphics[width=0.30\textwidth]{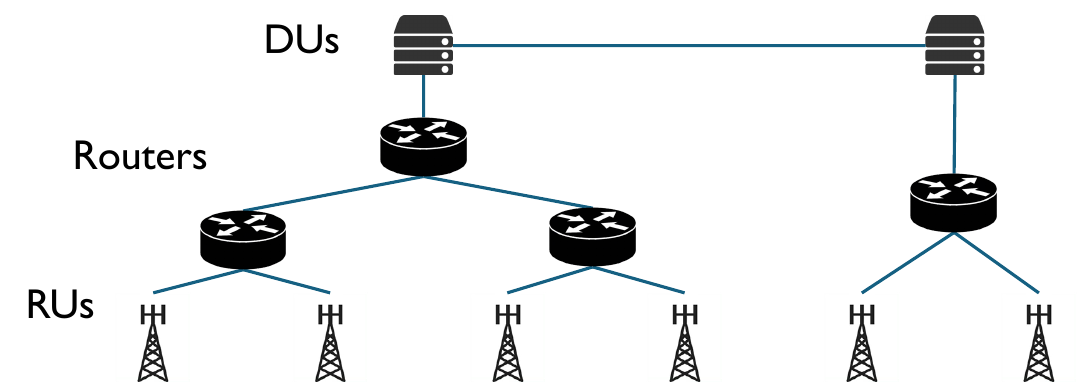}%
        \label{fig:force-tree}%
    }\hfill
    \subfloat[Folded Clos topology]{%
        \includegraphics[width=0.30\textwidth]{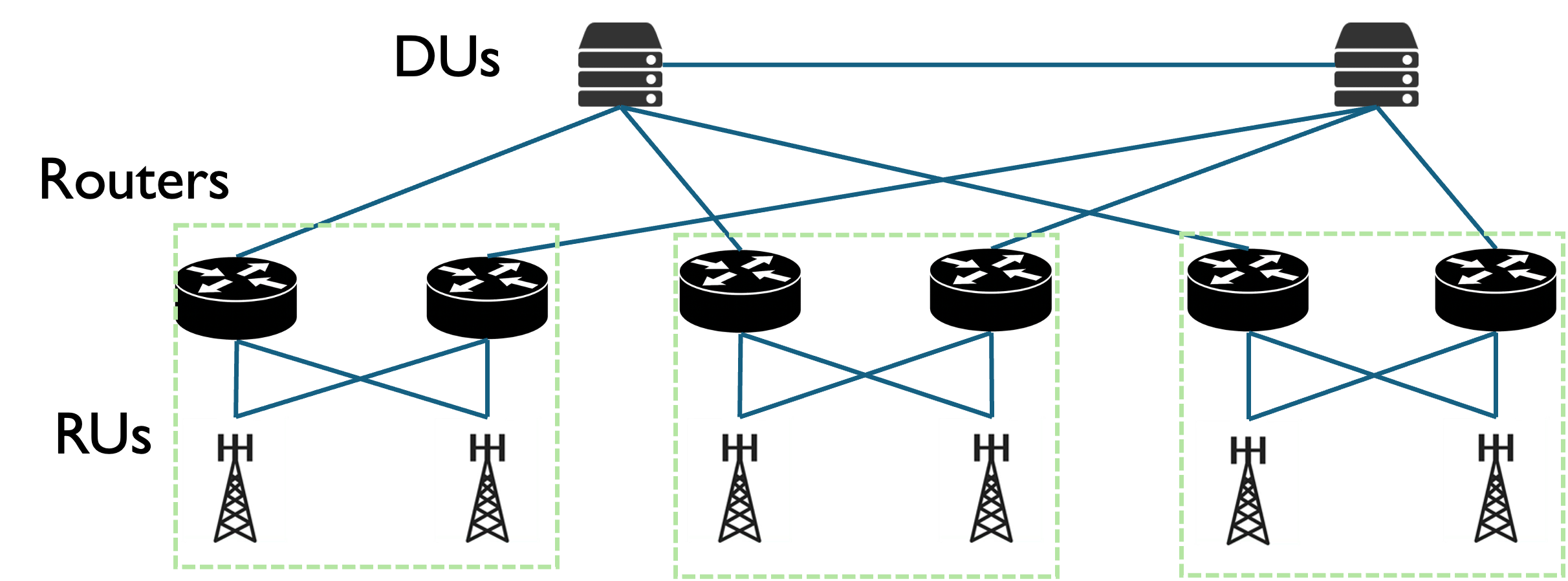}%
        \label{fig:fct}%
    }%
    \caption{A schematic view of topologies considered in this paper. 
    In Figure~\ref{fig:fct}, the dotted green boxes show the groups containing the~RUs.}
    \label{fig:Topologies}
\end{figure*}


Our graph model, based on the system model, considers four types of nodes: (1) \UEs (2) \RUs, (3) routers and (4) \DUs, as part of an underlying graph $G$.  We consider three different topology types for the underlying graph.

\noindent \textbf{Complete Topology.} Here, we consider that \RUs are directly connected to all \DUs.
We consider this type of graph as a baseline. See an example in Figure~\ref{fig:complete-topology}. 




\noindent \textbf{Tree Topology.} The previously discussed PON networks are known to be easily deployable. In their basic shape, they have a tree like structure. In our work, we adhere to the following steps to build a tree topology: 
Initially, we connect all \DUs together, considering a ring topology. Next, based on the number of \DUs, number of \RUs and the varying maximum degrees, we determine how routers can be put in different levels.  In the end, we connect the layers in a tree like topology. See an example 
in Figure~\ref{fig:force-tree}, in which $\Delta^\DU$ = $2$ and $\Delta^R$ = $3$.

\noindent \textbf{Clos Topology.} 
A \emph{Clos Topology} is a graph in which nodes are put into three layers, and each layer is connected to the layer before and after it, in a more dense manner compared to a Tree structure (for more general description of Clos topologies, refer to~\cite{Weight2}).  In our context, layers of a Clos topology come in the form of the \RUs, \DUs and an intermediate layer of routers, respectively from top to bottom (at the bottom layer, \RUs are connected to \UEs as usual). 
We partition the first layer into groups of \RUs with size $p$.  Then for each group, we add $q=\frac{Q}{N}$ routers, and connect these routers to all the corresponding \RUs in that group. 
Lastly, we connect the $i$-th \DU to the $i$-th router in each group. See an example of Clos topology shown in Figure~\ref{fig:fct}.

\subsection{Cluster placement algorithms}
We consider the following heuristics for the assignment of each user-centric cluster to one of the DUs. We propose two simple (but fast) demand-oblivious and two demand-aware assignment algorithms.

\noindent  \textbf{\textsc{Random}.} This algorithm assigns UE cluster to \DUs uniformly at random.

\noindent  \textbf{\textsc{Fixed}.} This algorithm divides the \UEs equally among \DUs, hoping that demand of \UEs would also be distributed equally among \DUs. This algorithm is designed to work well in scenarios where \UEs are distributed uniformly. 

\noindent  \textbf{\textsc{Aware}.} This algorithm first computes the shortest path from each \UE to all \DUs, and sorts \UEs by the sum of their uplink demands. Then, it goes over the sorted list of \UEs, and selects a valid \DU (that has not reached its capacity)  that minimizes increase in link utilization. 

\noindent  \textbf{\textsc{Hybrid}.} This is an extension of the \textsc{Aware} algorithm, that among the remaining valid \DUs computes a \emph{hybrid score} for each path based on both its length and current link utilization.




\section{Empirical Evaluation}
\label{sec:Eval}



\subsection{Setup}
 
\noindent \textbf{Input generation.} We consider the number of \RUs in the range $L =[20, 100]$, increasing in steps of $10$, where the number of RU antennas is $M = 10$. For our first set of experiments, we have $K = 3.5 L$ \UEs, so that $K$ is slightly less than $LM/2$. With user load $K \approx LM/2$, the network operates at optimal SE \cite{gottsch2023fairness}. 
To study how different user loads and spatial distributions (hotspots instead of uniform placement) impact the fronthaul traffic, in a second set of experiments we consider $K = 5 L$ and $K = 6.5 L$. We consider 10 instances of each scenario, and report the average demands.

The UE locations are generated randomly for each instance of $L$, such that in each instance we have different user-centric clusters and fronthaul requirements. In particular, let us consider UE $k$ for a given instance. In the UL, each RU $\ell \in \Cc_k$ must send $(1 - { {\gamma_{\rm DL}}})(T - \tau_p) W \fhratekUL$ bits through the fronthaul to the DU running the cluster processor of $\Cc_k$, where $W$ is the physical layer channel bandwidth (i.e., the used spectrum on the PHY). 
In the DL, the DU must send 
${\gamma_{\rm DL}}(T - \tau_p)W \fhratekDL$
bits to each RU $\ell \in \Cc_k$. 
The input is thus given by the RAN graph 
and the corresponding UL and DL fronthaul requirements $\{ (1 - { {\gamma_{\rm DL}}})(T - \tau_p) W \fhratekUL, { {\gamma_{\rm DL}}}(T - \tau_p)W \fhratekDL \}$
for each RU-UE association $(\ell, k)$, forming the demand matrix $\demmatrix$. 
As the UE locations are generated randomly for each instance, the 
inputs of each scenario are different.

We notice that the maximum channel bandwidth is $W = 100$ MHz in frequency range 1 (i.e., sub-6 GHz systems with carrier frequency $f_c = [0.41 , 7.125 ]$ GHz) and $W = 400$ MHz in frequency range 2 (i.e., mmWave systems with  $f_c = [24.25, 52.6]$  GHz) \cite{3gpp38104}. In this work, we consider a sub-6 GHz system. At a higher carrier frequency such as in mmWave systems, more spectrum is available to support wider channel bandwidths, which in turn allows more symbols to be transmitted per second.
For mmWave systems, it is thus expected that the fronthaul requirements will be larger due to the wider channel bandwidth.

\noindent \textbf{Simulation parameters.} We set the maximum outgoing degrees of network elements $\Delta^{\DU} = \Delta^{\RU} =  \Delta^{\Router} = 10$. We then consider up to $10$ \DUs, and set $Q = 1.5 L$. Furthermore, the maximum number of user-centric clusters that can be supported by each \DU is $\beta = \frac{K}{10}$.


\begin{figure}[t]
    \centering
    \subfloat[]{%
        \includegraphics[width=0.22\textwidth,clip]{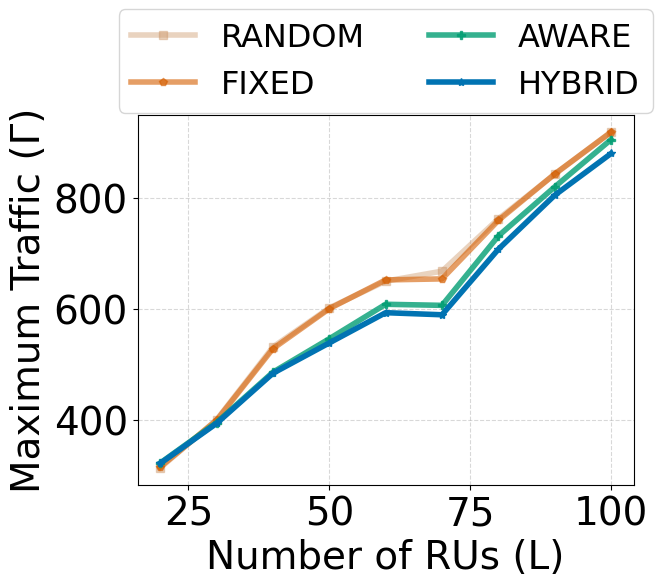}%
        \label{fig:alg-perforamnce-T}%
    }\hfill
    \subfloat[]{%
        \includegraphics[width=0.22\textwidth,clip]{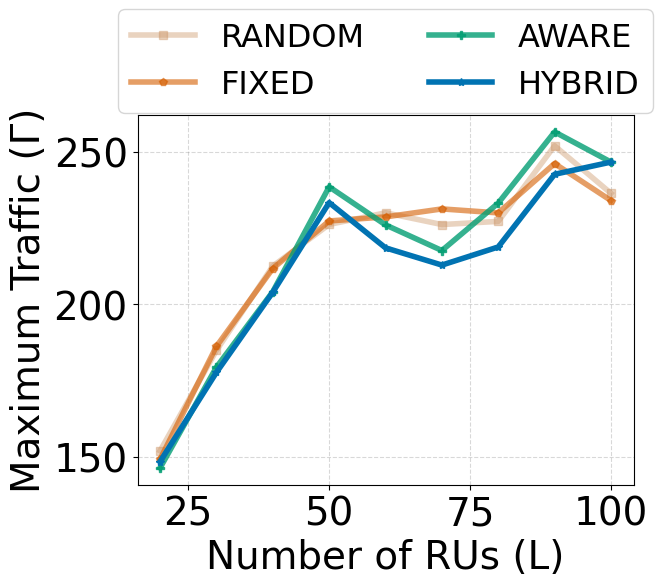}%
        \label{fig:alg-perforamnce-C}%
    }%
    \caption{Maximum traffic vs. number of RUs. The underlying topology is
    (a) tree and (b) Clos, respectively.}
    \label{fig:alg-perforamnce-2}
\end{figure}

\subsection{Results}
We simulate our algorithms in python~$3.10$, benefitting from the Matplotlib~\cite{Matplotlib} library. The following results report the maximum per link traffic $\Gamma$ in Gbps. The possible minor irregularities are due to the realistic nature of inputs, even when averaging over several iterations. This first set of experiments used $K = 3.5 L$ for all instances with a uniform user location distribution over the area.

\noindent \textbf{Effect of Algorithms.} First, we focus on the performance of suggested algorithms, in terms of maximum link traffic. Our simulation results show that the maximum link traffic increases linearly with an increase in the number of \RUs. In particular, we observe that on average, \textsc{aware} and \textsc{hybrid} algorithms outperform the \textsc{random} and \textsc{Fixed} algorithms. The difference between different algorithms can be up to 20\%, as Figs. ~\ref{fig:alg-perforamnce-T} and ~\ref{fig:alg-perforamnce-C} show for the Tree and Clos topology, respectively.

\begin{figure}[t]
    \centering
    \subfloat[]{%
        \includegraphics[width=0.24\textwidth,clip]{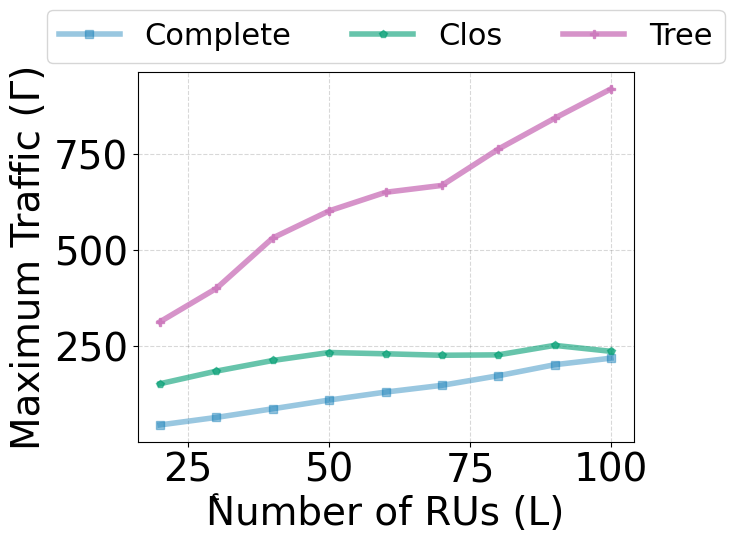}%
        \label{fig:topology-perforamnce}%
    }\hfill
    \subfloat[]{%
    \includegraphics[width=0.22\textwidth,clip]{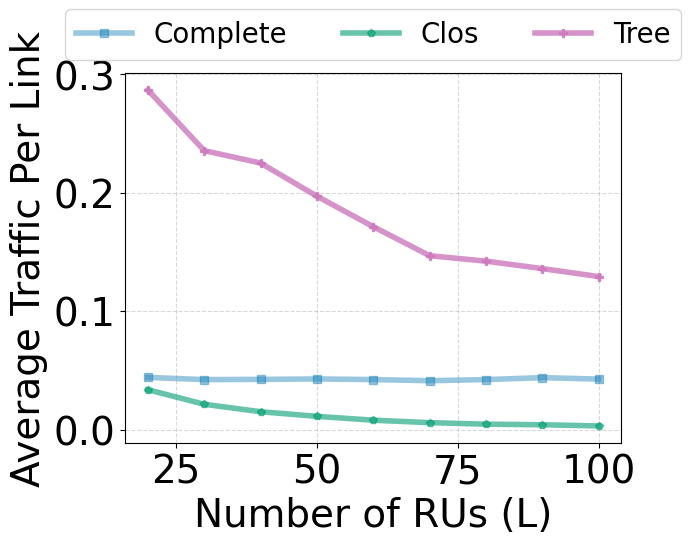}%
        \label{fig:topology-perforamnce-per-link}%
    }%
    \caption{Maximum traffic for various topologies~(a) and average maximum link traffic~(b) for a varying number of RUs. 
    The underlying algorithm in both figures is the \textsc{hybrid} algorithm.}
    \label{fig:topology-performance-all}
\end{figure}

\noindent \textbf{Effect of Topology.} Secondly, we compare the difference in maximum link traffic, considering various topologies. Here, our simulation results indicate that the Clos topology can significantly decrease the traffic in the fronthaul network compared to more traditional tree topologies by around 75\%.
    We observe that the Clos topology's performance gets much closer to the ideal complete topology, as the number of \RUs grows. Fig.~\ref{fig:topology-perforamnce} shows a comparison of different topologies with increasing number of \RUs.
    Furthermore, we report that in terms of average link traffic, the Clos topology is the best suited among the topologies that we have considered, see Fig.~\ref{fig:topology-perforamnce-per-link}. 

\begin{figure}[t]
    \centering
    \subfloat[]{%
    \includegraphics[width=0.24\textwidth,trim={0 0 0 0},clip]{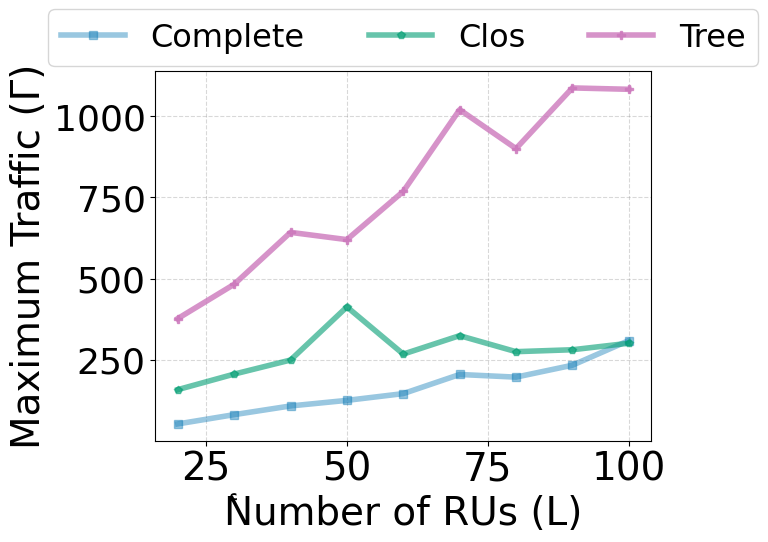}%
        \label{fig:Hotspot5K}%
    }\hfill
    \subfloat[]{%
    \includegraphics[width=0.24\textwidth,trim={0 0 0 0},clip]{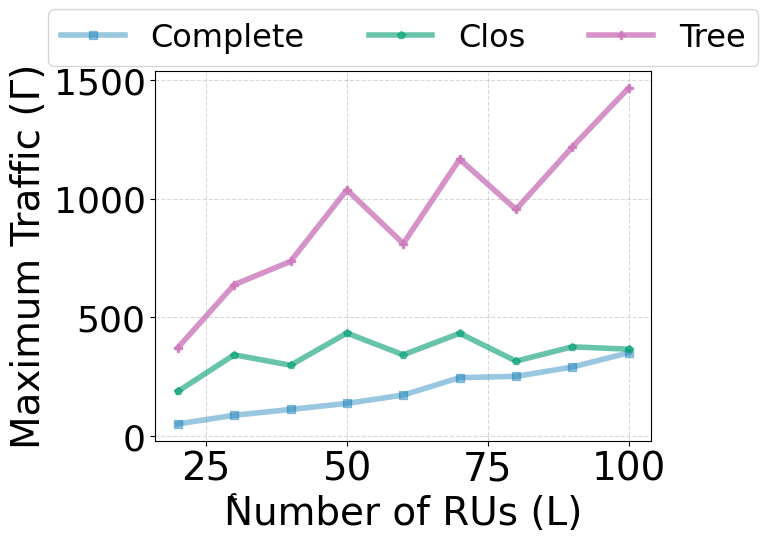}%
        \label{fig:Hotspot65K}%
    }%
    \caption{Maximum traffic for a hotspot distribution with (a)~$K = 5L$ and (b)~$K = 6.5L$.}
    \label{fig:Hotspot}
\end{figure}

For our second set of experiments, we modified the average number of UEs per RU. In addition, we also distributed the UEs in different patterns to emulate "hotspots" in the network area where many users congregate. 
The results for $K = 5 L$ are shown in Fig. \ref{fig:Hotspot5K} and for $K = 6.5 L$ in Fig. \ref{fig:Hotspot65K}. We observe similar behavior for the different fronthaul topologies as in the scenario with uniform distribution of user locations.

\subsection{Discussion}
Our results confirm our assumptions with regard to the feasibility of tree-like topologies for the fronthaul of future cell-free RANs. With growing number of RUs, they quickly reach a point where existing optical transport technologies can no longer support them. 


Keeping with the assumption made of using 100 GBit links, even for  small amounts of RUs, e.g. $L=20$, the required capacities already reach over 300 GBit, tripling the available capacities. Furthermore, they breach even the currently highest available point-to-point capacities of 800 GBit starting at around 90 RUs.
Higher carrier frequencies, such as mmWave or sub-THz, would push the demands up by at least another magnitude, with some researchers predicting data rates of up to 1 TBit per RU \cite{corre2019sub}. This highlights that it is not the radio/physical layer that is posing bandwidth and capacity limitations in cell-free MIMO, but instead the fronthaul network. It is thus up to network designers and planners to come up with new architectures and approaches to overcome these challenges. 
We showed that fast heuristics can have an impact on the link capacity requirements. While it is not yet  clear which algorithm is ideal, traffic reductions of up to 20\% are observable in certain scenarios.

Finally, we aim to find topologies that improve performance while allowing for feasible deployments. Using complete graphs as a baseline, we show that Clos topologies offer almost on par performance, especially as the number of RUs grows. As shown, while overall maximum traffic approaches that of an ideal complete graph, the average link traffic actually drops below, indicating even better load balancing of traffic across different links. We also note that our results hold for various distributions of UE locations in the mobile network area. 
\section{Conclusion}
\label{sec:conclusion}
Our findings show that cell-free 6G networks will not be able to live up to their full potential if the current state-of-the-art passive optical networks with tree topologies are used for the fronthaul. Instead, rethinking the architecture and topologies of fronthaul networks will be required to deal with the new demands and requirements imposed by the physical layer. It is up to the network community to design topologies in such a way that they offer a good trade-off between cost, complexity and efficiency. We showed another example of Clos topologies versatility here, outside of their current use cases in datacenter, enterprise networks, etc.

\section*{Acknowledgment} 
The authors acknowledge the financial support by the Federal Ministry of Research, Technology and Space of Germany in the programme of “Souverän. Digital. Vernetzt.”  Projects 6G-RIC, xG-RIC and xG-NOVA, project identification numbers 16KISK030, 16KIS2434, and 16KISS003.
\bibliographystyle{IEEEtran}
\bibliography{refs-CF-UC}

\end{document}